\documentclass[aps,10pt,preprintnumbers,final]{revtex4}
\usepackage{graphicx}
\begin{document}
\title{A low-temperature device architecture for the statistical study of electrical characteristics of 256 quantum devices}

\author{H. Al-Taie$^{1,2}$, L. W. Smith$^2$, B. Xu$^2$, P. See$^3$, J. P. Griffiths$^2$, H. E. Beere$^2$, G. A. C. Jones$^2$, D. A. Ritchie$^2$, M. J. Kelly$^{1,2}$ and C. G. Smith$^2$}

\address{$ˆ1$ Centre for Advanced Photonics and Electronics, Electrical Engineering Division, Department of Engineering, 9 J. J. Thomson Avenue, University of Cambridge, Cambridge CB3 0FA, United Kingdom}
\address{$ˆ2$ Cavendish Laboratory, Department of Physics, University of Cambridge, J. J. Thomson Avenue, Cambridge CB3 0HE, United Kingdom}
\address{$ˆ3$ National Physical Laboratory, Hampton Road, Teddington, Middlesex TW11 0LW, United Kingdom}

\begin{abstract}
Research in the field of low-temperature electronics is limited by the small number of electrical contacts available on cryogenic set ups. This not only restricts the number of devices that can be fabricated, but also the device and circuit complexity. We present an on-chip multiplexing technique which significantly increases the number of devices locally measurable on a single chip, without the modification of existing fabrication or experimental set-ups. We demonstrate the operation of the multiplexer by performing electrical measurements of 256 quantum wires formed by split-gate devices using only 19 electrical contacts on a cryogenic set-up. The multiplexer allows the measurement of many devices and enables us to perform statistical analyses of various electrical features which exist in quantum wires. We use this architecture to investigate spatial variations of electrical characteristics, and reproducibility on two separate cooldowns. These statistical analyses are necessary to study device yield and manufacturability, in order for such devices to form the building blocks for the realisation of quantum integrated circuits. The multiplexer provides a scalable architecture which makes a whole series of further investigations into more complex devices possible.
\end{abstract}

\maketitle

Email: ha322@cam.ac.uk, lws22@cam.ac.uk\\

Cryogenic experimental set ups limit the complexity of low-temperature electronics by the small number of electrical wires available on a cryostat. This also restricts the number of devices that can be fabricated on a single chip, and multiple cooldowns of many samples are necessary for electrical characterisation of devices~\cite{Yang2009}. Not only is this extremely time consuming and costly, but variations between cooldowns may lead to misleading characterisation of electrical features due to changes in the electron density.

For devices to be considered for applications in spintronics~\cite{Awschalom2007} or quantum information processing~\cite{Loss1998}, one must consider reliability, yield and manufacturability/reproducibility. 
One device which has been identified as having great potential in these applications~\cite{Chen2012, Debray2009} is the split-gate transistor~\cite{Thornton1986}.
The split gate is an ideal component to investigate because it can be seen as a building block for more complex devices (such as laterally-defined quantum dots). It is also the simplest device to exhibit quantum phenomena; the quantisation of conductance in units of $2e^2/h$ as electrons become confined in a one-dimensional (1D) channel~\cite{vanWees1988,Wharam1988}. 

We presented the first study of a large number of individually-addressable, gate-defined quantum devices on a GaAs/AlGaAs heterostructure~\cite{Al-Taie2013}. Separate measurement of each device was enabled using an on-chip increase in number of electrical contacts: A quantum multiplexer. 
The multiplexer addresses an array of 256 geometrically-identical split gates. Since many devices can be measured during a single cooldown in a cryostat, a study of the fabrication and quantum yield can be performed~\cite{Al-Taie2013}. Additionally, statistical variations in quantum phenomena can be investigated~\cite{Smith2013}. The focus of this present study is the reproducibility of electrical characteristics, both from device-to-device and after thermal cycling.

Devices were fabricated on a modulation-doped GaAs/AlGaAs heterostructure, in which the two-dimensional electron gas (2DEG) forms $90$ nm below the surface of the wafer. 
Data are presented before and after illumination with a light emitting diode. The carrier density ($n$) and mobility ($\mu$) were measured to be $1.7\times10^{11}$ cm$^{-2}$ and $0.94\times10^6$ cm$^2$V$^{-1}$s$^{-1}$, respectively, before illumination. After illumination, $n=2.9\times10^{11}$ cm$^{-2}$ and $\mu=2.2\times10^6$ cm$^2$V$^{-1}$s$^{-1}$. Standard optical lithography was used to define the surface gates, and the split gates ($0.4$ $\mu$m long and $0.4$ $\mu$m wide), were patterned using electron-beam lithography. Two-terminal lock-in measurements were performed at 1.4 K using excitation voltages between 25 and 100 $\mu$V at 77 Hz.

\begin{figure}[t]
\centering
\includegraphics[width=28pc]{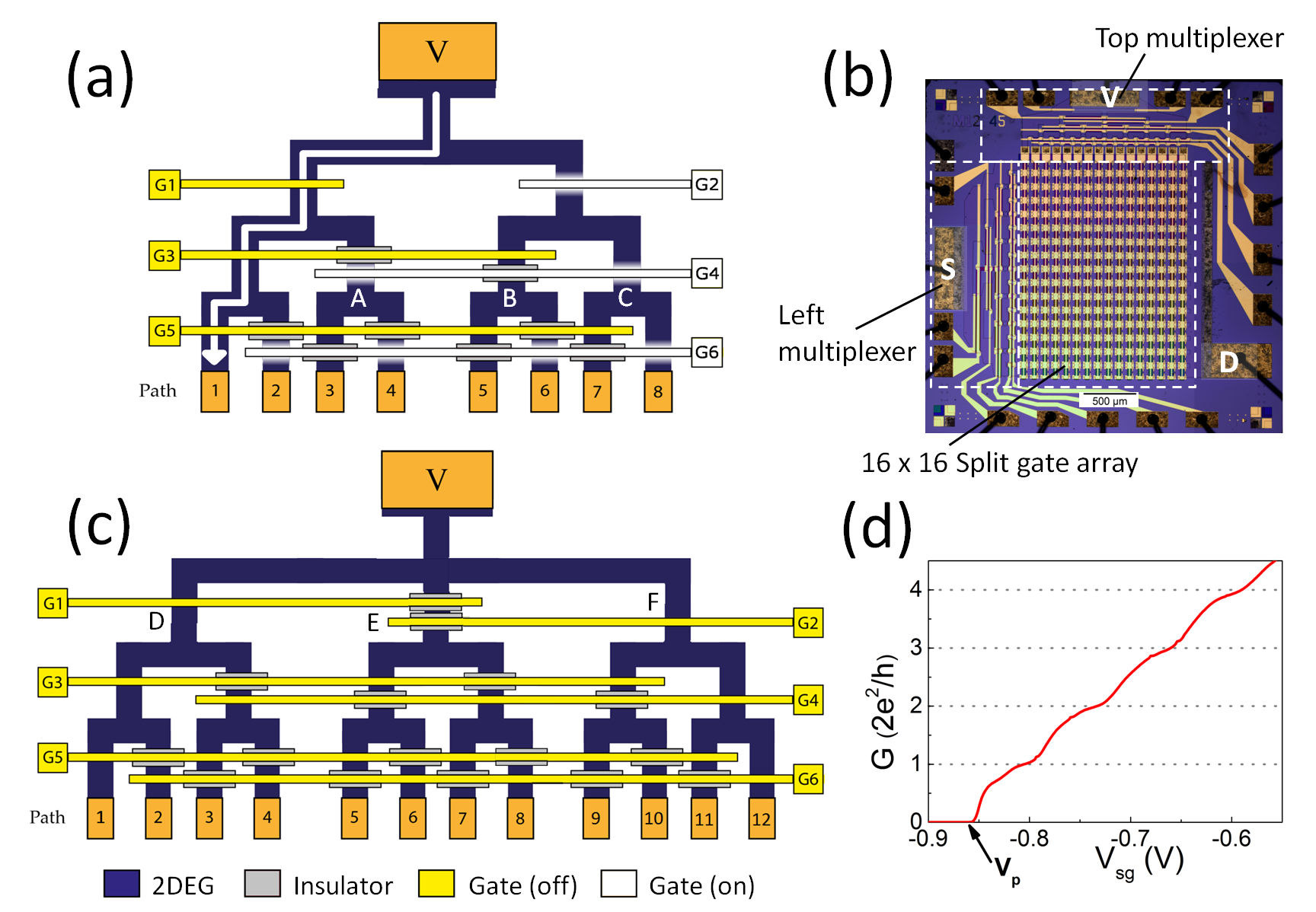}
\caption{\label{fig1}(a) Schematic diagram of a two-way quantum multiplexer. The mesa (blue) branches from input contact $V$ to eight separate output paths, labelled 1 - 8. Addressing gates are shown in yellow (off) and white (on), and insulating regions are shown in grey. Negative voltages applied to addressing gates $G2$, $G4$ and $G6$ route the input voltage along the path indicated by the arrow, to output 1. (b) Optical micrograph of the device, consisting of two multiplexers addressing an array of 256 split gates. The contacts for the source, drain and input-voltage are labelled S, D and V, respectively. (c) Schematic diagram of a three-way quantum multiplexer. (d) Conductance against $V_{sg}$ for an example split gate, in which $V_p$ is indicated by the arrow.}
\end{figure}

Figure~\ref{fig1}(a) illustrates the operation of the multiplexer. A raised `mesa' which confines the 2DEG branches from the contact labelled $V$, to which an input voltage ($V_{i}$) is applied. The output path through the multiplexer is determined by 6 `addressing gates' ($G1$ to $G6$). These gates cover multiple mesa-defined branches of the multiplexer; in Fig.~\ref{fig1}(a) gate $G4$ crosses the mesa at points $A$, $B$, and $C$. A photo-definable insulator (polyimide HD4104), is located  underneath the addressing gates at selected locations, for example at $B$. Therefore, a negative voltage applied to $G4$ is sufficient to deplete the 2DEG at $A$ and $C$, but not $B$. 
To direct $V_{i}$ along path 1 (as illustrated by the arrow), only gates $G2$, $G4$ and $G6$ are `on', i.e. a negative voltage is applied. The $V_i$ can be directed to any of the other output paths depending on which addressing gates are `on' or `off'. A voltage offset ($V_o$) must be maintained between $V_i$ and the addressing gates that are `on', where $V_o$ is the voltage which must be applied to each addressing gate to deplete electrons in the 2DEG when $V_i = 0$. 
Addressing gates which are `off' are held at the same voltage as $V_i$. 

Figure~\ref{fig1}(b) shows an optical micrograph of the chip, in which the source, drain, and contact to which the split-gate voltage ($V_{sg}$) is applied are labelled S, D, and V, respectively. 
The split gates are arranged in a rectangular array, and each device is measured individually, using two multiplexers (the left multiplexer selects the row, and the top multiplexer selects the column). The operation of the chip is described more fully in Ref.~\cite{Al-Taie2013}.

As a proof-of-concept, Fig.~\ref{fig1}(c) shows a schematic diagram of a three-way multiplexer, which gives a 50$\%$ increase in the number of output paths while maintaining the same number of electrical contacts (i.e. 6 addressing gates and 1 input contact). The operation is identical to the two-way multiplexer except that if the desired path is through either outer mesa channels $D$ or $F$, the voltage applied to addressing gates $G2$ or $G1$ must be large enough to deplete the 2DEG beneath the insulator at $E$. However, if the required path is through the central channel ($E$), a voltage is applied to both $G1$ and $G2$ such that the 2DEG is depleted at $D$ and $F$, but not at $E$.
This provides a scalable architecture that can be extended to address an even larger number of devices on chip. The number of output paths from the two-way ($N_{2mux}$) and three-way ($N_{3mux}$) multiplexers are given by
$N_{2mux}=2^{(n-1)/2}$ and $N_{3mux}=2^{(n-1)/2}+2^{(n-3)/2}$, respectively, where $n$ is the number of electrical contacts (including the addressing gates and input contact). In either design, adding two additional addressing gates doubles the number of output paths.

\begin{figure}[t]
\centering
\includegraphics[width=42pc]{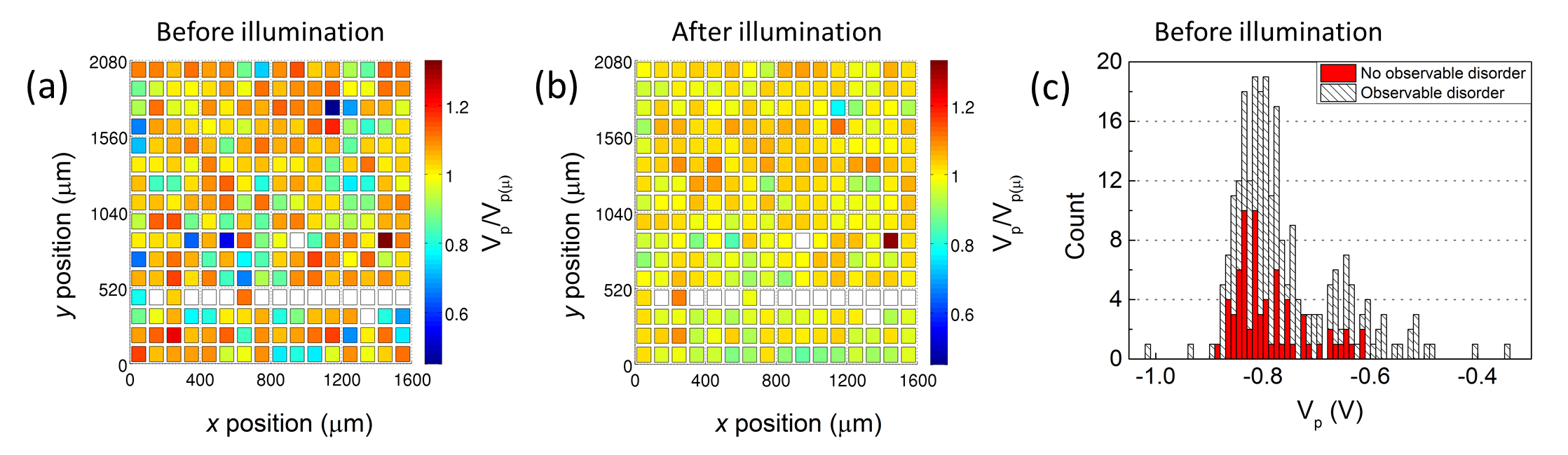}
\caption{\label{fig2} Colour scale showing the spatial distribution of V$_{p}$ normalised to V$_{p(\mu)}$ (a) before illumination and (b) after illumination, where each box represents a split gate in the array. Clear boxes indicate split gates which did not define a 1D channel, therefore V$_{p}$ could not be determined. 
(c) Histogram of $V_{p}$ before illumination, for a bin size of $10$ mV. The data are categorised according to whether disorder effects were or were not evident, represented by shaded and solid bars, respectively.}
\end{figure}

The conductance ($G$) through each split gate was measured individually, as a function of $V_{sg}$. Due to damage that occurred during fabrication, fifteen split gates did not define a 1D channel, leaving 241 usable devices. This gives a yield of $94\%$, which may be improved with careful manufacture.
Figure~\ref{fig1}(d) shows conductance $G$ as a function of $V_{sg}$ for an example split gate (before illumination). The pinch-off voltage ($V_{p}$) is defined as the voltage at which $G = 0$, marked by the arrow.

Figures~\ref{fig2}(a) and (b) show colour-scale plots of $V_p$ as a function of spatial distribution before and after illumination, respectively. The $V_{p}$ is normalised to the mean pinch-off voltage [$V_{p(\mu)}$], and the range of the colour bars in (a) and (b) are identical, to allow direct comparison.
Each box represents a split gate in the array, and clear boxes indicate the 15 split gates where $V_{p}$ could not be determined.
From Fig.~\ref{fig2}(b) it can be seen that the normalised variation of $V_p$ reduces after illumination.
Prior to illumination $V_{p(\mu)} = -0.76$ V, and standard deviation $\sigma = 97$ mV.
Following illumination, $V_{p(\mu)} = -2.86$ V (due to the increased $n$), and the absolute value of $\sigma$ increased to $144$ mV. 
However, there is a reduction in the coefficient of variation (defined as a dimensionless ratio of the standard deviation to the absolute of the mean), from 0.127 to 0.054, such that a higher $n$ appears to lead to greater uniformity of $V_p$.

Quantum phenomena in mesoscopic devices can be affected by disorder. Two main sources of disorder in GaAs HEMTs are scattering from ionised donors and ionised impurities. In 1D measurements, this can lead to deviations in the quantised conductance from $2e^{2}/h$, weakened or missing plateaux, and evidence of Coulomb blockade~\cite{Liang1997}.

Figure~\ref{fig2}(c) shows a histogram of $V_{p}$ (before illumination), for a bin size of $10$ mV. The data are categorised according to whether disorder effects were observed at any point on the conductance trace. Before illumination, 70 (171) devices did not (did) show evidence of disorder (in this measurement $T=1.4$ K, at a lower $T$ more disorder effects may become apparent). The range of $V_p$ is smaller where disorder was not observed ($\approx 0.3$ V), and the standard deviation $\sigma = 67.3$ mV compared to $104.7$ mV. The total distribution in Fig.~\ref{fig2}(c) is perhaps bi-modal, for which an explanation has yet to be determined.

Certain disorder effects (such as missing or weakened plateau), were much less evident after illumination, since the higher electron density results in better screening of impurities. However, after illumination almost all devices showed occasional length resonant-like features in the conductance, perhaps due to the stronger 1D confining potential~\cite{Kirczenow1989}.

\begin{figure}[t]
\includegraphics[width=42pc]{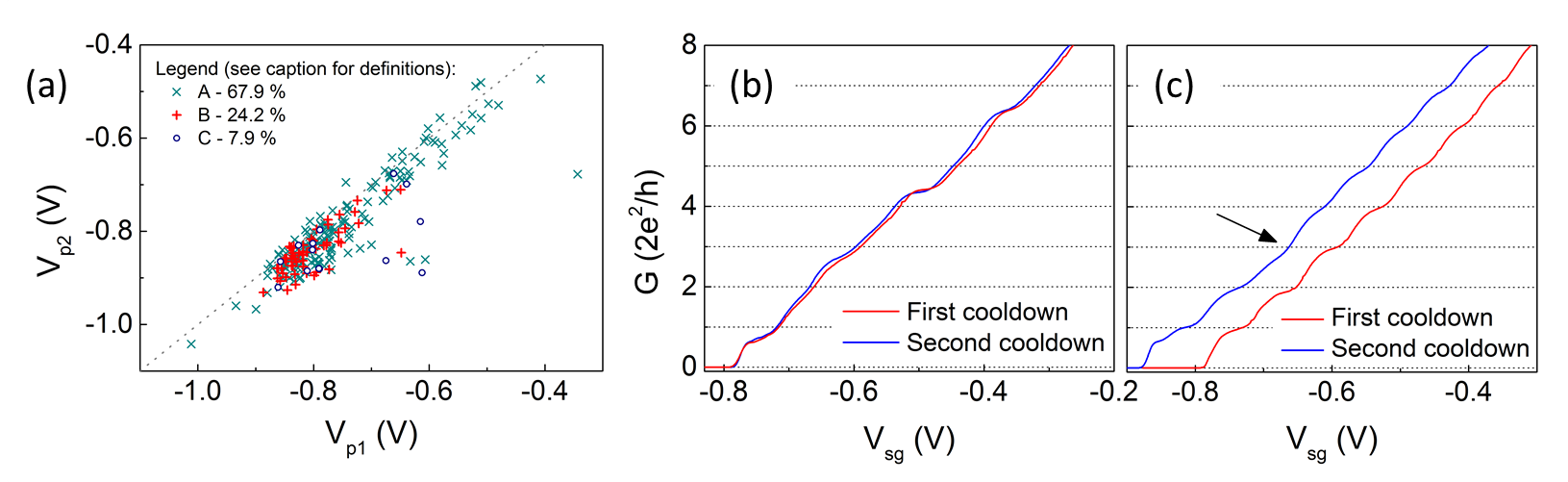}
\caption{\label{fig3}(a) Scatter plot of $V_p$ from each split gate on two separate cooldowns; $V_{p1}$ and $V_{p2}$. Categories A, B and C refer to whether the devices showed evidence of disorder on both, neither, or only one cooldown. As a guide, the dotted line shows $V_{p1} = V_{p2}$.
(b), (c) Conductance as a function of $V_{sg}$ for two example split gates. In each panel the conductance traces from both cooldowns are shown. In (c) the arrow marks a weakened plateau, evidence of disorder on the second cooldown.}
\end{figure}

The sample was thermally cycled and measured on a second cooldown.
Figure~\ref{fig3}(a) shows a scatter plot of $V_{p2}$ against $V_{p1}$, where $V_{p1}$ ($V_{p2}$) is $V_p$ from the first (second) cooldown. An additional split gate did not define a 1D channel on the second cooldown, leaving a total of 240 devices.
The data are categorised according to whether disorder was observed on both cooldowns (67.9$\%$), on neither cooldown (24.2$\%$), or only on one cooldown (7.9$\%$). 
Therefore, whether disorder effects were or were not observed did not change for 92.1$\%$ for devices. 
The Pearson-product moment correlation coefficient $r$ quantifies the degree of correlation between two variables. For Fig.~\ref{fig3}(a) $r=0.89$ (including all data), indicating a very strong correlation. 
The mean change in pinch-off voltage $\delta V_{p(\mu)}=39.2$ mV, which may be related to a change in $n_{2D}$ on successive cooldowns.
A large $\delta V_{p}$ ($>0.15$ V), occurred for only 7 devices, irrespective of whether disorder effects were observed.
Since $V_{p}$ is dependent on electron density, a large $\delta V_{p}$ may reflect local variations in $n$ due to a redistribution of ionised donors in the dopant layer. 

By comparing conductance data from successive cooldowns it may be possible to distinguish between likely sources of disorder. Figure~\ref{fig3}(b) shows $G$ as a function of $V_{sg}$  for a device in which the conductance characteristics are highly disordered. Despite the disorder, the conductance is remarkably similar between cooldowns.
The strength of the effect, and the reproducibility on successive cooldowns suggests that an ionised impurity exists near the 1D channel. 

Figure~\ref{fig3}(c) shows $G$ against $V_{sg}$ for another device, where the arrow indicates a weakened plateau on the second cooldown.
Such cooldown-dependent effects may be related to remote ionised donors, since the distribution of donors that are ionised varies on thermal cycling.
Additionally, the effect is relatively weak, consistent with the origin of the disorder being further away from the 1D channel (i.e. separated from the 2DEG by a 40 nm wide spacer layer).

In summary: the reproducibility of electrical characteristics in 1D devices has been investigated, as a function of spatial location and after thermal cycling. This was made possible using a scalable on-chip multiplexer, which allowed 256 split gates to be individually measured during a single cooldown.
This approach can be used to determine limiting factors on device reproducibility, and identify where to direct efforts for improvement.
In the present study, disorder, variations in the density, and the re-distribution of dopant ions on the second cooldown were all limiting factors.
Improvements can be made by fabricating devices on a higher quality GaAs/AlGaAs heterostructure (quantified by a greater $\mu$), which will likely result in smaller variations of $V_p$ between devices, and fewer observable disorder effects. Additionally, dopants can be removed entirely by fabricating devices on undoped heterostructures~\cite{Kane1993}. 
It may also be instructive to investigate how variations in density can be reduced by controlling the cooling process, and by illuminating the sample during the cooldown.
This is an initial study, which has presented a framework for considering reproducibility in quantum devices. It can be used to test the suitability of devices for applications in quantum computing and spintronics.

\section*{Acknowledgements}
This work was supported by the Engineering and Physical Sciences Research Council Grant No. EP/I014268/1. The authors would like to thank C. J. B. Ford, I. Farrer and F. Sfigakis for invaluable discussions, and R. D. Hall for electron-beam exposure.

\end{document}